\documentclass{article}
\usepackage{fullpage}

\usepackage{amsmath}
\usepackage{amsfonts}
\usepackage{setspace}
\usepackage{enumerate}
\usepackage{graphicx}

\usepackage{breakcites}
\usepackage[round]{natbib} 

\DeclareUnicodeCharacter{0301}{\'{e}}

\usepackage[bookmarks]{hyperref} 
\hypersetup{
    colorlinks=true,
    linkcolor=black,
    citecolor=cyan,
    urlcolor=black,
}
\usepackage{bookmark}

\newcommand{\jargon}[3][]{\footnote{#2: #3}}

\usepackage{authblk}

\title{Linking fast and slow: the case for generative models}

\author[1]{Johan Medrano}
\author[1]{Karl J. Friston}
\author[1]{Peter Zeidman}

\affil[1]{The Wellcome Centre for Human Neuroimaging, UCL Queen Square Institute of Neurology, London, WC1N 3AR, UK}

\date{} 

\begin{document}
\maketitle

\begin{abstract}
    A pervasive challenge in neuroscience is testing whether neuronal connectivity changes over time due to specific causes, such as stimuli, events, or clinical interventions. Recent hardware innovations and falling data storage costs enable longer, more naturalistic neuronal recordings. The implicit opportunity for understanding the self-organised brain calls for new analysis methods that link temporal scales: from the order of milliseconds over which neuronal dynamics evolve, to the order of minutes, days or even years over which experimental observations unfold. This review article demonstrates how hierarchical generative models and Bayesian inference help to characterise neuronal activity across different time scales. Crucially, these methods go beyond describing statistical associations among observations and enable inference about underlying mechanisms. We offer an overview of fundamental concepts in state-space modeling and suggest a taxonomy for these methods. Additionally, we introduce key mathematical principles that underscore a separation of temporal scales, such as the slaving principle, and review Bayesian methods that are being used to test hypotheses about the brain with multi-scale data. We hope that this review will serve as a useful primer for experimental and computational neuroscientists on the state of the art and current directions of travel in the complex systems modelling literature. 
\end{abstract}


\section{Understanding the brain: a multiscale problem}
    Phenomena that span multiple temporal scales are ubiquitous in brain research \citep{breakspear2017dynamic, dangelo2022quest}.  For instance, an objective of epilepsy research is to understand the genesis of seizures by analysing the itinerancy of cortical circuitry parameters over seconds to minutes \citep{Rosch2018, jirsa2014nature, jirsa2023personalised,courtiol2020dynamical,depannemaecker2021modeling, lopez2022personalizable,lisi2018markov,depannemaecker2023phenomenological,ponce2015resting}. In naturalistic experiments, researchers study the dynamics of the freely-behaving brain in exchange with its environment, on the scale of minutes to hours \citep{Lee2021,echeverria2022transient,Shain2020,demirtacs2019distinct,meer2020movie}. On a larger time scale, studying the progression of Alzheimer's disease requires accounting for processes evolving over months, years or even decades \citep{Wang2020,Graham2021}. In these examples, a challenge is to  understand how slow and fast components interact; in other words, how slowly changing variables shape the evolution of rapidly changing ones, and how the latter reciprocally influence the former \citep{ellis2012top}. 

    This calls for a statistical framework for evaluating hypotheses when empirical observations span multiple temporal scales, or when the hypotheses are framed on a different time scale from that of the observations. Quantitatively comparing the evidence for hypotheses amounts to building generative models and comparing their ability to predict the data, in terms of model evidence (a.k.a., marginal likelihood). Modelling multiple temporal scales affords the potential to better predict future observations. 
    
     This review is based on the overarching idea that evaluating hypotheses with multiscale time series can be systematically addressed by constructing hierarchical models, where each level of the hierarchy accounts for a different temporal scale. Separating temporal scales can be rigorously justified and — by pairing such models with Bayesian procedures – they become useful tools for investigating the causes of observed data.  

    How can hierarchical models be used to capture changes in neuronal connectivity at different temporal scales, and what is the mathematical justification for their use? How to evaluate the evidence for competing hierarchical models, in order to address neuroscientific hypotheses or make modelling decisions (e.g., which temporal scales should be modelled)? These questions are the focus of this review. In the three sections that follow, we first introduce key concepts for modelling the dynamics of time series data at a single timescale. We then detail principles and strategies for separating timescales within mathematical models. Third, we introduce Bayesian statistical methods for evaluating the quality of multi-scale models, before concluding with an empirical example in the field of epilepsy research. 
     
\section{Modeling time series}
    To test hypotheses about the temporal structure of our observations, we need to model their generating process. Clearly, the form of the model depends on the nature of the data, e.g., whether it is continuous or discrete, and on the assumed nature of the generating process, e.g., deterministic or stochastic. This section introduces several key concepts in time series modelling, before considering a unified view of the different approaches that are suitable for neuroimaging. These concepts provide the foundation for linking temporal scales, detailed in the remainder of the paper. 
    
    \subsection{State-space models}       
        When observations come from a continuous set of values, the modelling problem lends itself to the apparatus of dynamical systems. In particular, state-space models capture the evolution over time of some latent or unobserved variables called states, $\mathbf{x}$, which give rise to observations $\mathbf{y}$. Using this framework, which was established in the 1960s in the field of control theory \citep{kalman1960general}, a broad class of dynamical systems can be modelled using equations of the form
        \begin{align}
            \label{eq:dynsys}
            \begin{cases}
                \frac{d}{dt}\mathbf{x}(t) &= f(\mathbf{x}(t), \mathbf{u}(t), \theta)\\
                \mathbf{y}(t) &= g(\mathbf{x}(t), \mathbf{u}(t), \theta) + \omega(t)
            \end{cases}
        \end{align}
        where all variables can be vectors. The first equation is the evolution equation or state equation. At time $t$, the velocity of the system's states $\frac{d}{dt}\mathbf{x}(t)$ depends on the states themselves $\mathbf{x}(t)$, some inputs $\mathbf{u}(t)$ and some parameters $\theta$, which can be viewed as states that change infinitely slowly. The states are observed through the observations $\mathbf{y}(t)$ produced by the observation equation. The observations might be corrupted by some additive measurement noise $\omega(t)$. 

\begin{figure}[t!]
            \centering
            \centerline{\includegraphics[width=1.\textwidth]{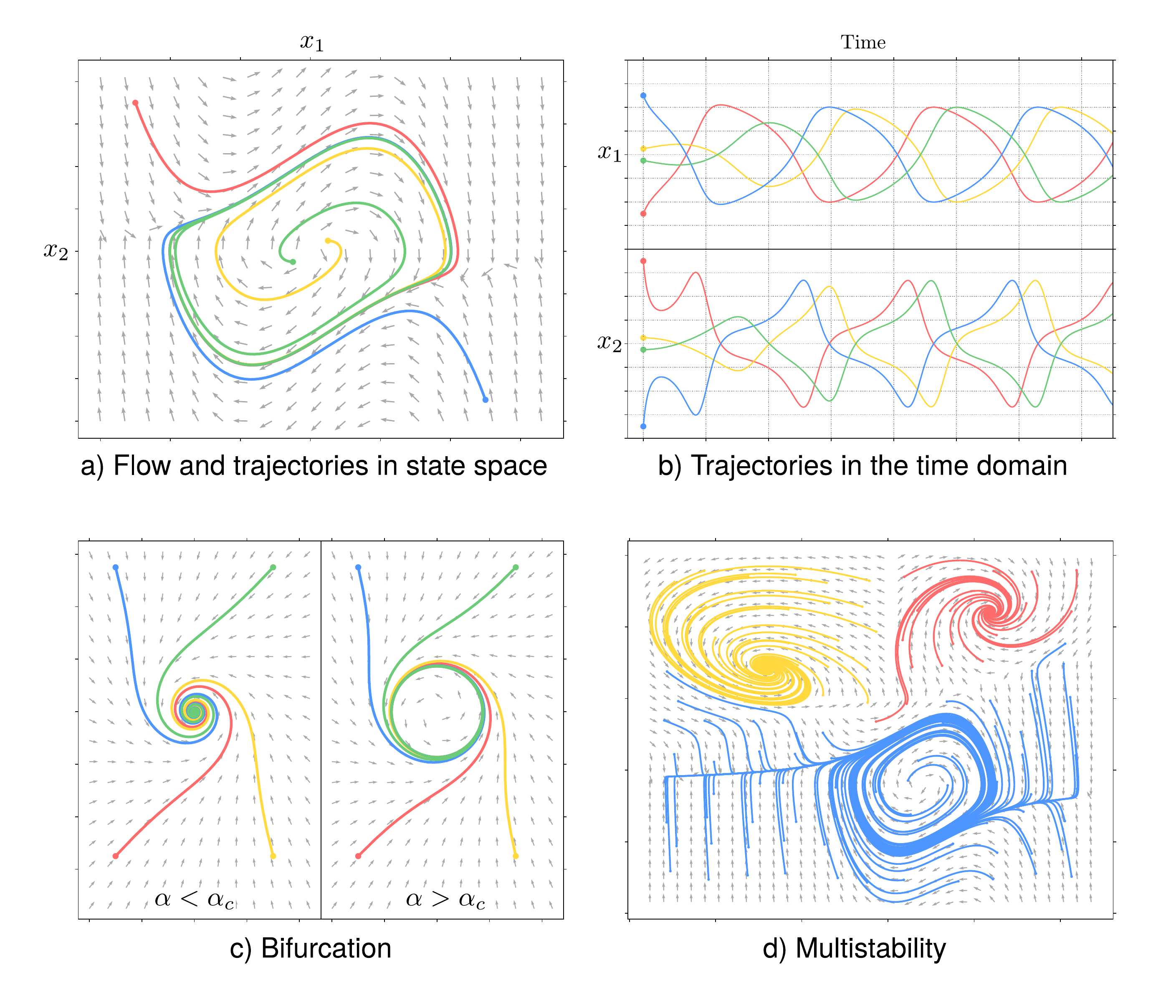}}
            \caption{Important concepts of the theory of dynamical systems. a) An example of flow in state-space (grey arrows), governing the evolution of trajectories (coloured curves) from different initial states (coloured circles). b) The corresponding trajectories in the time domain for both $x_1$ and $x_2$ axes. c) An example of bifurcation of the flow: trajectories converge towards a fixed point of state space when the bifurcation parameter $\alpha$ is below a critical value $\alpha_c$, and towards a limit cycle when the bifurcation parameter is above the critical value (Andronov-Hopf bifurcation). d) An example of a multistable system: the attractor to which the trajectory evolves  depends on the initial state, as indicated by the colours of the trajectories. }
            \label{fig:dynsys101}
        \end{figure}

         A state space is a geometric space, the axes of which are the states of the system (for example, the firing rates of different neuronal populations), as illustrated in \textbf{Figure~\ref{fig:dynsys101}}. Given some inputs and parameters, the first line of Eq~\eqref{eq:dynsys} attributes a velocity vector at each point of state space. The mapping from states to velocity is the flow of the system and plays an important role in understanding its evolution: every state trajectory evolves along the flow. The heterogeneity of the flow (intuitively, arrows pointing at each other) gives rise to attractors, i.e. dense sets of points (for instance, a single point, a circle, or a sphere) attracting trajectories in their basin of attraction. Attractors define the steady state behaviour of the system: any trajectory in the basin of attraction evolves towards --- and remains in --- the attractor. 
         
         A multistable dynamical system has multiple attractors, thus its steady state depends on the attraction basin in which the system is initialized. Additionally, changes in inputs or parameters can cause bifurcations, i.e. changes in the flow such that the topology of the attractor changes, for instance from a fixed point to a limit cycle \citep{guckenheimer1983nonlinear}. Bifurcations and multistability appear only in nonlinear dynamical systems and play a critical role in modelling biological systems \citep{haken1978synergetics, haken1985theoretical, kelso2012multistability, deco2012ongoing, tognoli2014metastable, breakspear2017dynamic, jirsa2020structured, ponce2023critical}.

            When the evolution is purely deterministic as in Eq~\eqref{eq:dynsys}, one can represent the output of a dynamical system as a function of past inputs, without reference to internal states. In other words, the output signal can be derived from the inputs, without having to solve an \textbf{initial value problem}\jargon{\textbf{Initial value problem}}{The problem of finding solutions to a differential equation with given initial conditions.}, i.e., to integrate the equations through time. For linear systems, the dependence of outputs on past inputs is captured by a linear \textbf{response function}\jargon[36pt]{\textbf{Response function}}{A function that describes the output or behaviour of a system in response to a given input or stimulus.} which, convolved with an input time series, gives the output time series. For nonlinear systems, a series of higher-order response functions (a Volterra expansion) captures the dependence of outputs on past inputs \citep{fliess1983algebraic}. 
            
            An example application of response functions in functional MRI (fMRI) research is the use of a Volterra formulation to summarise the dynamics of the Blood-Oxygen Level Dependent (BOLD) response \citep{friston2000nonlinear}. In this case, the first-order Volterra kernel corresponds to the change in the BOLD response due to a brief stimulus, and the second-order Volterra kernel quantifies how the BOLD response to one stimulus depends on the time elapsed since the previous stimulus (i.e., non-linear effects of time). These Volterra kernels can be generated from the underlying state-space model. In practice, one is interested in finding the simplest model of the observations that can explain the most data (this is formally motivated thereafter). In fMRI, the relevant properties of nonlinear systems are well-captured by retaining the first terms of the Volterra expansion, which corresponds to approximating the flow of the nonlinear dynamical system with a bilinear \citep{friston2003dynamic, friston2000nonlinear} or second-order polynomial form \citep{stephan2008nonlinear}. 
            
            For modelling neuronal connectivity and dynamics, the response function of a dynamical system plays an important role, because it captures how the system transforms the frequency spectrum of the input signal. This is particularly helpful for modelling systems that are driven by random fluctuations, such as neural mass models, which capture the activity of populations of neurons. When driven by endogenous cortical noise, it is impossible to track the evolution of a neural mass model: despite having a deterministic evolution, their trajectories depend nonlinearly on a particular realisation of the input noise process which is unavailable to the experimenter. In that case, we can model the output spectrum of neural masses by applying their linearized response function to their input spectrum \citep{moran2007neural, moran2013neural, moran2011consistent, chen2008dynamic, symmonds2018ion, rosch2018nmda}. The use of a linear approximation can be motivated by the fact that coupled neural systems exhibit linear dynamics over short timescales \citep{heitmann2018putting}.  

            Dynamic Causal Modelling (DCM) \citep{friston2003dynamic, frassle2021tapas} and The Virtual Brain \citep{sanz2013virtual,jirsa2017virtual} are examples of  applying state space modelling to neuronal dynamics, and in particular for investigating \textbf{effective connectivity}\jargon{\textbf{Effective connectivity}}{The directed influence of neuronal populations on one another.}; namely, the effects of neuronal populations on one another. In this context, the states are continuous variables, and the evolution equation typically describes the change in neuronal activity over time, whereas the observation function generates the measurements one would expect to make, e.g., using electro-encephalography (EEG) or functional magnetic resonance imaging (fMRI).

        \subsection{A taxonomy of modelling frameworks}

            Forming a model necessitates two key decisions. First, whether the dynamics will be deterministic, as in Equation 1, or stochastic, meaning there will be some degree of randomness to the evolution of the states. Stochastic models are covered by the more general framework of  random dynamical systems \citep{arnold1995random}. The second decision is whether the states themselves will be discrete or continuous variables. Examples of  these different kinds of modelling frameworks are provided in \textbf{Figure~2}.
            
            A special case arises when the dynamics are stochastic, and the states are discrete values. This typically calls for Markov chains and related models \citep{fraser2008hidden}. Markov chains are stochastic processes that model probabilistic transitions between a finite collection of states. More precisely, Markov chains model the evolution of the states' distribution, i.e.,\ the probability of being in a certain state at a certain time, by means of state transition probabilities, which gives the probability of switching from a state to another. Markov chains fulfil the Markov property: knowing the current state distribution and the state transition probabilities is sufficient to determine the future state distribution. 
    
            Markov chains have the same role as the evolution equation of dynamical systems. As such, they can also be equipped with an observation function transforming abstract "states" into the expected distribution over observations. The resulting models are known as \textbf{Hidden Markov Models}\jargon{\textbf{Hidden Markov Model}}{A probabilistic model that describes a sequence of observable events generated by an underlying unobservable Markov chain.} (HMMs) and are popular devices in modelling timeseries \citep{rabiner1989tutorial, bishop2006pattern, vidaurre2016spectrally, baker2014fast, vidaurre2017brain, cabral2017cognitive}. Relevant examples of HMM applications include (i) analysing the switching dynamics of resting states networks in a large cohort \citep{vidaurre2018discovering}, (ii) evaluating how brain network dynamics are altered during natural movie watching \citep{meer2020movie} and (iii) identifying brain networks activated during replay \citep{higgins2021replay}. 

            Both stochastic and deterministic state-space models are routinely used in the DCM modelling framework, which is implemented in freely available software (SPM \citep{penny2011statistical}, TAPAS \citep{frassle2021tapas}). DCM pairs the specification of models with Bayesian system identification techniques, in order to estimate the evidence for alternative candidate models. Recent developments have included continuous state-space models of psychiatric disease progression \citep{friston2017computational}, whole-brain effective connectivity in resting state fMRI \citep{frassle2018generative, frassle2021regression}, as well as discrete state-space models of Covid-19 progression \citep{friston2020second, friston2022dynamic}. \textbf{Figure~2} includes examples of particular variants of DCM that handle different kinds of states and evolution equations.
    
            To summarize, time series can be modelled as resulting from the dynamics of latent states, where the dynamics may be deterministic or stochastic and where the states may be continuous or discrete. Despite the apparent heterogeneity of modelling methods, they are constructed from common components: an evolution equation, guiding the system's state through time, and an observation equation, producing the measured quantities. Response functions can then be derived from the state-space model, serving as a useful device for obtaining the output of a system from its input. The choice of approach is determined only by the nature of the states and that of the evolution equation, for the particular application domain.

            \begin{figure}
                \centering
                \includegraphics[width=\textwidth]{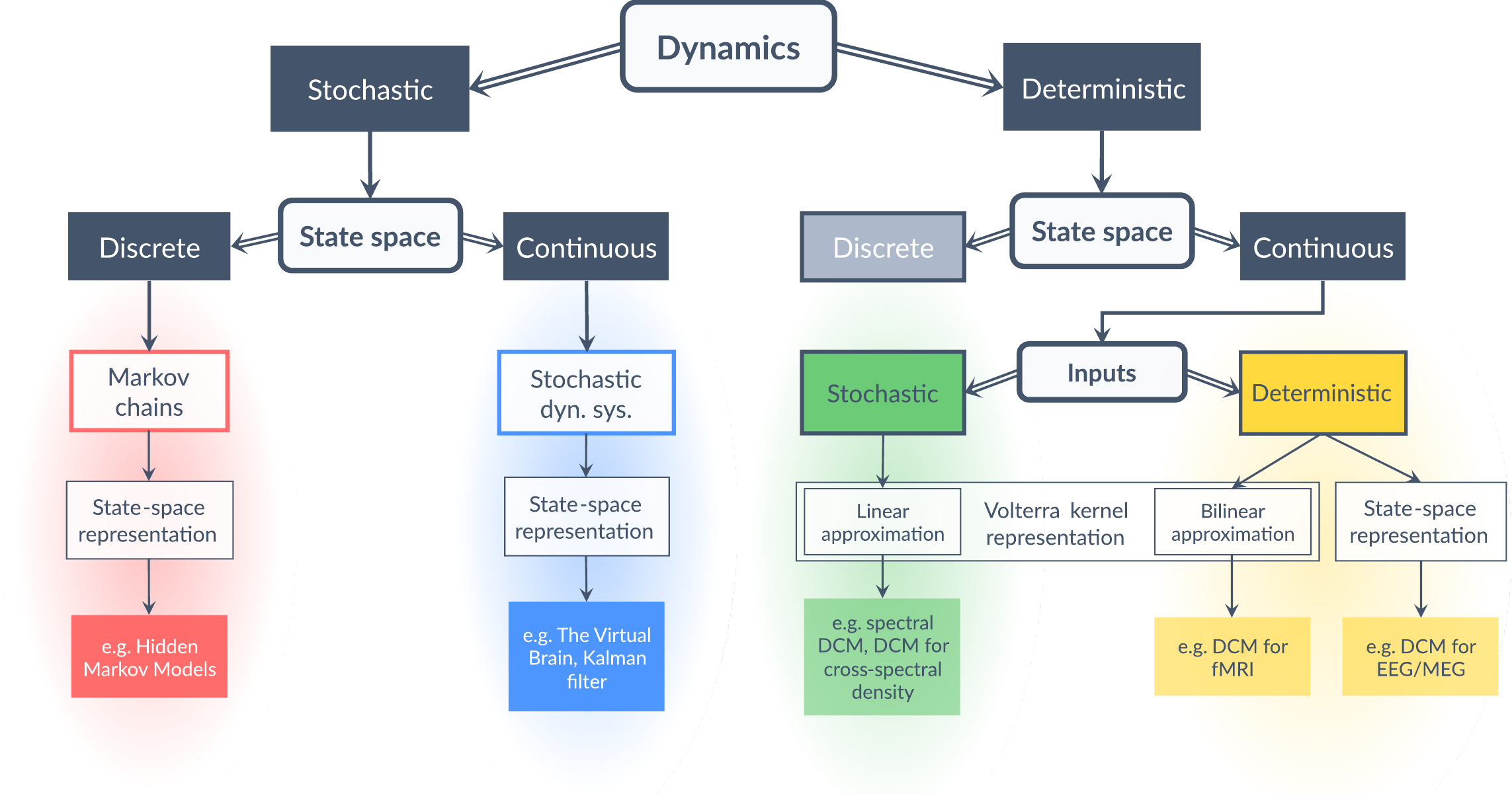}
                \caption{Taxonomy of the different modelling frameworks discussed here. The key factors guiding the selection of a particular framework are the nature of dynamics (discrete or continuous) and the nature of state space (stochastic or deterministic). In addition, the nature of the inputs (stochastic or deterministic) has relevance for continuous deterministic systems. In the particular case of continuous deterministic systems with stochastic inputs one can use a linear response function, which is the first-order term of the Volterra kernel representation of the system, to directly approximate the outputs from the inputs without reference to the states. Effectively, this implies that the dynamics do not need to be integrated over time, which greatly simplifies model inversion. In all other cases, model inversion entails tracking the states or their distribution through time. }
                \label{fig:taxo}
            \end{figure}
                     
\section{Separating timescales}
        The fundamental concepts in time series modelling reviewed above can be applied to situations where observations have a temporal structure at different temporal scales. This is the case when the duration of the observations is long enough to allow the slower temporal features to unfold --- and when the sampling rate is high enough to capture fast temporal features. In this case, modelling the evolution at different time scales becomes crucial, in order to infer the common underlying generative processes or mechanisms that give rise to the data.

    \subsection{Motivation}
        Considering the temporal evolution of time series at various scales is a natural approach for studying neuronal dynamics. One common practice involves examining power fluctuations in the frequency bands of EEG or MEG data. The power spectrum of a signal summarizes short-term signal changes, while the power of specific frequency bands (a..k.a., band power) reveals longer-term variations. By using a model with parameters governing the shape of the spectral density, such as the response function of a neural mass model, we can connect the rapid fluctuations in electrical activity over milliseconds to the slow evolution of these parameters over seconds or minutes.

        These models allow us to quantify how experimental manipulations relate to continuous changes in slow parameters. For instance, \citep{rosch2018nmda} demonstrated how the concentration of an epilepsy-inducing drug modulated the power spectra of neuronal activity by affecting certain synaptic rate constants. In the future, this type of model could prove valuable in characterising the dynamics of the self-inhibition of superficial pyramidal neurons, thought to encode prediction precision in predictive coding accounts of brain function, for instance during naturalistic experiences \citep{bastos2012canonical,kanai2015cerebral,adams2016dynamic}.
        
        However, building and analysing models with dynamics over different temporal scales can be challenging. Determining the appropriate temporal scale for each variable – and understanding its interaction with other variables – is not straightforward. In many cases, interactions among variables evolving at different timescales can result in circular causality, complicating the analysis and leading to complex multiscale dynamics. A key example of this is experience-dependent plasticity, where slow fluctuations in synaptic efficacy depend upon fast fluctuations in neuronal activity. At the same time, distributed neuronal activity depends upon synaptic efficacy. Fortunately, there are mathematical principles that can help simplify these complex models while still capturing the essential aspects of their dynamics. 
        
        This section proceeds in three parts. First, we introduce mathematical arguments, starting with foundational and abstract concepts and gradually transition to more practical applications. Second, we explore the application of these arguments to stochastic dynamics. Last, we discuss their relevance to models with slow discrete dynamics such as HMMs.
    
    \subsection{Mathematical perspectives}
        In dynamical systems, different temporal scales appear when some states evolve quickly relative to others. That is, if each state is conceptualized as forming a dimension of a state space (as we saw in \textbf{Figure 1a}), then different temporal scales emerge when the flow governing the evolution of trajectories is stronger in some directions than in others. The components of the dynamics in the stronger directions of the flow converge rapidly towards their steady state, with negligible displacements along the weaker directions of the flow. From the perspective of fast components, slow components may be considered as, effectively, static. Reciprocally, on the temporal scale at which slow components evolve, the fast components instantaneously reach their steady state. Inherent to the concept of separation of temporal scales is the reduction of the number of dynamical degrees of freedoms: over long periods of time, the macroscopic behaviour of the system can be described by the evolution of slow variables \citep{kuramoto2019concept}. This natural separation of temporal scales is crucial for modelling high-dimensional systems such as the brain (see \textbf{Figure~3}).

        \begin{figure}[t!]
            \centering
            \includegraphics[width=\textwidth]{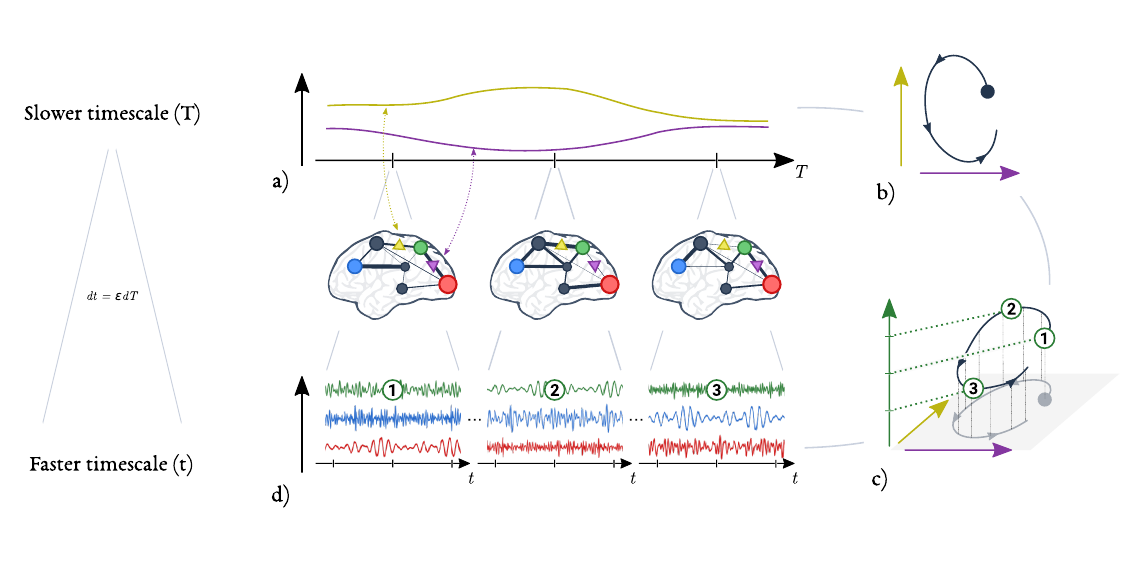}
            \caption{Multiscale dynamics of brain signals: mapping slow and fast variables. (a) Slow quantities in the brain, such as synaptic efficacy between regions, exhibit large time constants and evolve slowly over time. The evolution of two slow variables are illustrated here, as yellow and purple lines. (b) The evolution of slow variables can also be represented by dynamics in a slow state space. (c) Importantly, for every location in the slow state space (horizontal axes), there is a corresponding mode of fast dynamics (vertical axis). Three modes are depicted here, numbered 1-3. These fast dynamics give rise to rapid brain signals, such as field potentials in pyramidal neurons, as depicted in (d). The mathematical relationship between the slow and the fast timescale is given by ${dt} = \varepsilon{dT}$ ($\varepsilon \ll 1$); in words, the dynamics at faster scale $t$ unfolds over a fraction ($\varepsilon$) of the slower scale $T$. In summary, the brain is understood to navigate slowly (a) through a repertoire of fast stable dynamics (d). Crucially, the slow variables are directly linked to the dynamics of the fast variables (c). Similarly, changes in the fast variables' dynamics can be attributed to changes in the slow variables. Therefore, modelling the complex dynamics of multiscale dynamical systems can be simplified by focusing on the dynamics of the slow variables and the mapping from slow to fast variables.  }
            \label{fig:brain-signals}
        \end{figure}

        Mathematically, the separation of temporal scales in a nonlinear dynamical system can be proved rigorously. The centre manifold theorem (see \textbf{Figure~4}) is used to show the exponential decay of the fast components towards a manifold, i.e. a generalised surface that is aligned with the weak directions of the flow \citep{carr2012applications}. When the centre manifold theorem applies, one can consider that the dynamics are separated into fast and slow components evolving on different temporal scales. However, the centre manifold theorem is a local result; in other words, it can only be used to justify a separation of temporal scales in the vicinity of a fixed point. Recently, interest has been drawn to inertial manifolds. Similarly to centre manifolds, inertial manifolds are surfaces that attract trajectories exponentially quickly, but are not constrained to be local. However, in contrast to centre manifolds, there is no generic method to prove the existence of an inertial manifold, which needs to be assessed on a case by case basis. 
        
            \begin{figure}[t]
                \centering
                \includegraphics[width=\textwidth]{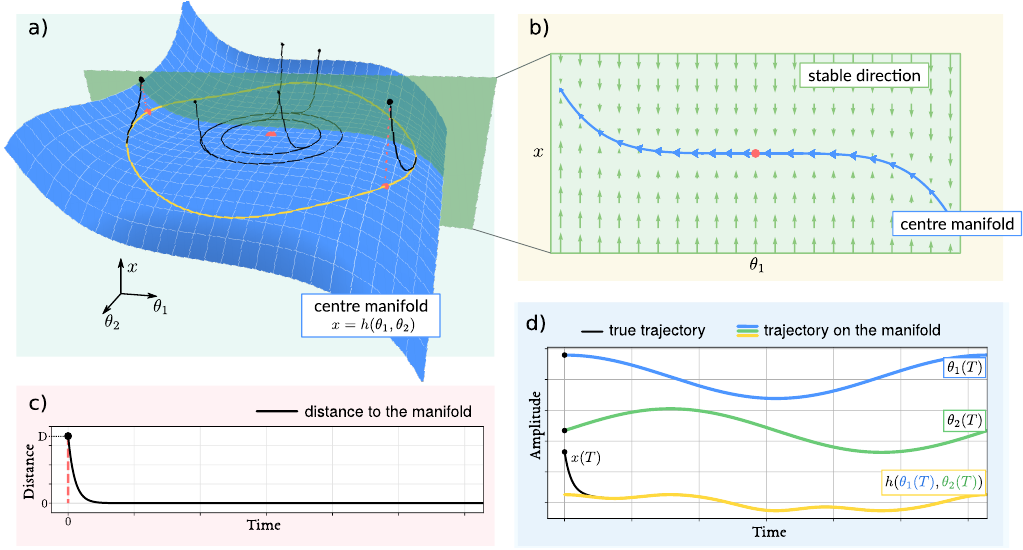}
                \caption{Illustration of the centre manifold theorem with a three-dimensional dynamical system. \textbf{a)} The three-dimensional state-space of the system. The blue surface is the centre manifold, and gives a height $x = h(\theta_1, \theta_2)$ to each point of the $(\theta_1, \theta_2)$ plane. Trajectories initialized away from the manifold converge  rapidly towards the surface (black curves). This is due to the presence of a strong flow orthogonal to the centre manifold, as show in \textbf{b)} for a section of state space. The strong flow (green arrows) converges towards the centre manifold (blue curve). The flow parallel to the manifold (blue arrows) is weaker by orders of magnitude. Hence, trajectories quickly collapse to the centre manifold before evolving alongside it. This is reflected in the exponential decay of the distance to the manifold, as shown in \textbf{c)}. Hence, the $x$-component of the trajectory is well approximated by a static function from the location on the $(\theta_1, \theta_2)$ plane \textbf{(d)}. This can motivate an adiabatic approximation: as the rapidly changing $x$ component of the trajectory can be approximated by the mapping $h(\theta_1, \theta_2)$, we may consider $x$ as a spurious dimension of the system and restrict our description to the evolution on the $(\theta_1, \theta_2)$ plane; in other words, we can approximate the fast vanishing states by a fixed mapping from the slow states.}
                \label{fig:cmt}
            \end{figure}
            
        Manifolds give a practical way to construct multiscale dynamical systems: one can specify an equation for the (inertial) manifold, mapping from slow to fast states, and prescribe strong normal (perpendicular) flow to make fast states quickly collapse onto the manifold. The framework of Structured Flows on Manifolds (SFM) adopts this procedure. SFM explains how slow low-dimensional variables, such as movement parameters, could be instantiated by the collective behaviour of a large number of fast variables, such as networks of neurons \citep{pillai2017symmetry, jirsa2019grand, Jirsa2022}. In SFM, the fast components of the flow drive neuronal activity to rapidly evolve into synchronization modes (functional modes) that correspond to points on the manifold. Simultaneously, the flow on the manifold progressively modifies the functional modes, thereby generating slowly changing behavioural variables.
        
        Frameworks such as SFM are important for brain research because they give a mechanistic account of brain activity and its relationship to behaviour \citep{mcintosh2019hidden,jirsa2022entropy}. Interestingly, this type of theoretical modelling approach aligns well with the needs of empirical work on intracranial recordings of neuronal activity. Typically, an objective is to understand how movement variables are encoded by stable low-dimensional dynamics of neuronal populations in the motor cortex \citep{churchland2012neural,saxena2022motor,lara2018different,langdon2023unifying,gallego2018cortical,chaudhuri2019intrinsic}. Combining more advanced modelling frameworks --- together with the statistical machinery described later --- can provide valuable tools to show on how empirical recordings support different theoretical perspectives on motor control. 

        We can also go one step further with the separation of temporal scales and neglect components perpendicular to the manifold when considering the evolution of slow variables (as illustrated in \textbf{Figure~4}). This separation is formally known as the \textbf{adiabatic approximation}\jargon{\textbf{Adiabatic approximation}}{Neglecting rapidly changing variables in comparison to slowly changing ones.}: the fast variables are replaced by their steady-state solution, leaving only the differential equations of the slow variables \citep{haken1978synergetics,friston2011network}. The adiabatic approximation is particularly appealing for multiscale systems as it allows one to replace references to the fast variables in the slow dynamics by a fixed mapping from slow variables, thereby finessing the problem of circular causality. 
        
        The mathematical arguments of slow-fast decomposition originate from theoretical physics and have been used in works on self-organisation and \textbf{synergetics}\jargon{\textbf{Synergetics}}{ The study of complex systems and their emergent properties, examining interactions between parts to understand their holistic dynamics.} \citep{haken1977synergetics, haken1978synergetics}. Haken and colleagues framed the decomposition of temporal scales in terms of a ``slaving principle'', where slow collective parameters, the order parameters, govern the dynamics of fast individual variables. This was elegantly used to construct the Haken-Kelso-Bunz (HKB) model of synchronisation during rhythmic finger movements, which provides a simple yet representative example of the slaving principle underpinning the slow-fast decomposition \citep{haken1985theoretical}. They found that when participants moved their index fingers at a fast speed, those initially moving in opposite phases suddenly synchronized. By using the slaving principle, they summarised the complex behaviour of the coupled finger dynamics through a single slow collective variable (the phase difference), which served as an order parameter. Their model shows that when the speed of finger movements exceeds a critical frequency, a bifurcation occurs in the order parameter: the antiphase movements become unstable, and the fingers synchronize. This celebrated example illustrates the power of the slaving principle in simplifying intricate multiscale phenomena. 

        To summarize, a dynamical system evolving on two different timescales can be described by a weak flow on a manifold, i.e.\ a surface within state space, and a component normal to the manifold whose amplitude decays exponentially with time. This fact can be used, not only to assess the existence of a separation of temporal scales in a given system, but also to construct a multiscale dynamical system by prescribing a manifold, a flow on the manifold, and a strong convergent flow normal to the manifold. This is the approach taken by the SFM framework. Multiscale dynamical systems can be further simplified by discarding the normal component, a step known as the adiabatic approximation. The adiabatic approximation has a particular advantage when dealing with dynamical systems driven by noisy inputs. 
        
    \subsection{Stochastic dynamics and conditioning}
        We have discussed separating timescales in continuous deterministic dynamical systems. However, the dynamics may be influenced by noise, and it is useful to understand how random fluctuations enter a dynamical system with multiple timescales. 

        A key mathematical result is that the centre manifold theory and the adiabatic approximation apply to stochastic systems \citep{knobloch1983bifurcations, boxler1989stochastic}. Naturally, the timescale separation takes a different form than in the deterministic case: we have to describe the evolution of the state \textit{probability distribution} over time. When the centre manifold theorem applies, the state distribution factorizes into a marginal, time-dependent distribution of slow modes $p(r,t)$ and a conditional, time-independent distribution of fast components  $p(s|r)$. In other words, the distribution of the fast modes is fully determined by the slow modes.

        The application of the centre manifold theorem to stochastic systems has important consequences. The distribution of fast modes is conditioned on the slow modes. For multiscale stochastic dynamical systems, this conditioning separates the different temporal scales. Thus, a generative model of multiscale time series naturally has a hierarchical structure, where the hierarchical depth accommodates the multiple fine- or coarse- graining of the systems temporal structure.  

    \subsection{Discrete systems and the slaving principle}            
        The slaving principle describes the general idea that slow quantities determine the evolution of fast quantities. For the particular case of continuous dynamical systems, the principle is evinced by the centre manifold theorem and the adiabatic approximation. However, the slaving principle is more general and also integrates common applications of discrete switching models such as HMMs. Here, we show how discrete systems can be reconciled with the overarching slaving principle. 
      
        When studying brain signals, it appears that brain activity is confined to modes or states of activity. These modes are defined by specific patterns of activity across the brain, which can be described through functional or effective connectivity between different brain regions. Notably, these stable modes of activity are activated in a sequential manner, and analysing the sequences of activation can provide valuable insights into the overall dynamics of the brain on a large scale. Analysing brain network dynamics -- especially in resting state -- has proved useful, for instance in pain research, in epilepsy research  \citep{jirsa2017virtual, rosch2018network} and in Alzheimer's research  \citep{kuang2019concise, nunez2021abnormal, smailovic2019eeg}. 
        
        A recent trend in neuroimaging involves employing HMMs to capture the sequential activation of the fast stable modes \citep{vidaurre2016spectrally, baker2014fast, vidaurre2017brain, cabral2017cognitive}. With electrophysiological data, the overall evolution of brain-wide neuronal activity is represented by both the cross-spectral coherence, which captures continuous short-term brain activity, and a Markov chain that models the switching dynamics over longer time scales \citep{vidaurre2016spectrally, o2018dynamics, vidaurre2018discovering}. The use of HMMs is licensed by the fact that the transitions between states happen rapidly, whereas the duration spent in each state is sufficiently long to enable the rapid activity to reach its steady state \citep{rabinovich2008transient, deco2012ongoing, rabinovich2018discrete, zarghami2020dynamic, ponce2015resting}. Thus, the switching dynamics of neuronal connectivity in HMMs adheres to the slaving principle: the slow Markov chain governs the dynamics of the fast brain signals, as captured by the cross-spectral coherence.

        To summarize, time series at different temporal scales can be separated and individually modelled. Separating temporal scales implies committing to the slaving principle, i.e.\ the idea that slow quantities govern the dynamics of fast quantities. The mathematical arguments underpinning the slaving principle are, for now, restricted to continuous systems; however, the principle can be intuitively applied to systems with switching dynamics such as HMMs.  Naturally, this perspective shows that a model of multiscale time series has a hierarchical structure, with each layer accounting for a different temporal scale. As a result, we can model the power spectra of EEG or MEG signals using a fast neural mass model alongside a slow model for parameter dynamics. Similarly, the dynamics of a group of neurons can be captured by a fast spike rate model along with a slow model for population dynamics. The next section will focus on how to utilize hierarchical models to test hypotheses with empirical data.
    
\section{Evaluating hypotheses with hierarchical models of multiscale time series}

    A mathematical model, of the sort described above, specifies a hypothesis about how the measured data were generated. Arbitrating between hypotheses entails specifying a suitable set of models, and having the statistical machinery in place to compare their evidence. This section sets out recent developments in Bayesian statistical methods, which are proving useful in neuroimaging and related fields that deal with multiscale time series data and dynamic models.

    \subsection{Statistical methodology and the case for modelling}
        In Bayesian statistics, the key quantity summarising the quality of a model is the log model evidence, also called the marginal likelihood, $\ln{p(y|m)}$. This is the log of the probability of having observed the data $y$ given the model $m$. Two or more models can then be compared in terms of their relative log evidence, which is called the log \textbf{Bayes factor}\jargon{\textbf{Bayes factor}}{A statistical measure that quantifies the evidence supporting one hypothesis, as represented by a model, over another.}: $\ln{B}=\ln{p(y|m_1)} - \ln{p(y|m_2)}$ \citep{kass1995bayes}.  This procedure is called Bayesian model comparison.
        
        During model development, Bayesian model comparison provides a principled framework for evaluating modelling decisions. Then, when testing hypotheses in an empirical setting, Bayesian model comparison serves as the basis for evaluating how well each model explains the data. This widespread use of the log evidence is justified because it has a number of attractive properties. For instance, it can be decomposed into the difference between the accuracy and complexity of a model, so it can be used to identify the simplest explanation for the data that maximizes the explained variance. Furthermore, through a generalization of the Neyman-Pearson lemma \citep{neyman1933ix,fowlie2021neyman}, it can be shown that the Bayes factor maximizes statistical power.
        
        Generative models that describe how neuronal dynamics give rise to observed neuroimaging data typically contain many parameters. These may include coupling strengths between brain regions or the rate of depolarization of a neuronal membrane. Prior knowledge about model parameters are generally available and the probability distribution over the parameters is refined by observing the data. Bayesian inversion refers to the process of transforming a prior probability distribution over the parameters into a \textbf{posterior distribution}\jargon[-33pt]{\textbf{Posterior distribution}}{The updated probability distribution of a variable after considering new data.} through observing the data. This depends on calculating or estimating the model evidence, which as stated above, quantifies how well the model explains the data, once any uncertainty about the parameters has been accounted for.
        
        Calculating the posterior distribution is generally intractable and two categories of methods exist to approximate it. The first approximates the posterior directly using numerical sampling. With the second category of methods --- variational Bayes --- one first specifies the functional form of an approximate posterior and then minimize a statistical difference between the true and approximate posterior. 
        
        Variational methods are preferred when the objective is to evaluate hypotheses by comparing models. Although sampling methods can be used to compute any kind of complex, multimodal posterior probability distribution, there is no standardized method to reliably compute the model evidence. In contrast, variational Bayes methods approximate the log model evidence using a quantity known as the variational free energy or an Evidence Lower Bound (ELBO). An optimization algorithm is applied, which searches for a setting of the model parameters that maximize the free energy. When maximized, the free energy approaches the log evidence. The quality of the approximation depends on the model complexity and becomes exact for general linear models. The variational Bayes treatment of stochastic and deterministic nonlinear dynamical systems has a long history of use in neuroimaging \citep{daunizeau2009variational,friston2003dynamic}.

    \subsection{Inverting hierarchical models}
        We have established that evaluating hypotheses amounts to comparing models in terms of the ratio of model evidence, and that models of multiscale time series have a hierarchical structure. This raises the following question: how to invert hierarchical models and estimate their model evidence? 
    
        To recap, each level of a hierarchical model has parameters that are constrained by the level above. For example, with neuronal recordings, a neural mass model may be used at the first level, with parameters that encode synaptic connection strengths. Slow changes in these synaptic parameters over a longer time period (e.g., seconds or minutes) are described by the second level of the model, with parameters that govern the slow dynamics.
        
        The Bayesian treatment of hierarchical models is particular. Because it is generally hard to prescribe \textit{a priori} the probability distribution of the parameters at arbitrary levels of the hierarchy, one is required to use \textbf{empirical priors}\jargon{\textbf{Empirical priors}}{Prior distributions estimated from the empirical distribution of data or lower-level posteriors.}. The posterior distributions at each level are constrained by the priors of the level above. However, using an empirical Bayes approach with hierarchical models creates a circular dependency: empirical priors must be estimated from the posterior distribution of levels below, which in turn depends on empirical priors of levels above. An established solution to the circularity problem is to use an iterative procedure and alternately update our estimates of empirical priors and posteriors \citep{efron1973stein,kass1989approximate}. 
       
       A good example of empirical Bayes if found in Bayesian approaches to group studies. Subject-level data are modelled with parameters that are constrained by a group-level prior distribution. Each subject's model is inverted using the group-level prior to produce a posterior estimate per subject. The prior group-level distribution is then refined from the collection of subject-level posterior estimates. The refined group-level empirical prior is used to produce new subject-level posterior estimates, which can be used to refine the empirical prior, etc. 

        \textbf{Bayesian model reduction}\jargon{\textbf{Bayesian model reduction}}{ A statistical technique based on Bayesian inference that simplifies complex models by eliminating irrelevant variables and retaining important ones.} is a simpler approach that avoids the need to iterate between levels of the hierarchical model and is used extensively in the context of neuroimaging \citep{friston2016bayesian}. The free energy --- applied here to approximate the log evidence of the entire hierarchical model --- can be separated into a partial free-energy for each hierarchical level. Each partial free-energy quantifies the complexity of that level and its accuracy at predicting the posterior density of the level below. Crucially, the free-energy of a level can be optimized regardless of higher levels. Thus, one can approximate the posterior of the first level using the empirical data, then approximate the posterior of the second level using the posterior of the first level, and repeat this procedure upwards in the hierarchy. This provides a straightforward analysis procedure for group studies in neuroimaging, which begins with modelling each participant's data individually (first level analysis), and then conveying the free energy and posteriors to the group level (second level analysis).

       The Bayesian model reduction approach to hierarchical models echoes the separation of temporal scales \citep{Jafarian2021}. In hierarchical models of multiscale time series, each level represents a different temporal scale, with higher levels representing longer timescales. With a genuine empirical Bayes procedure, one would iterate over different timescales, which would make the problem computationally demanding. However, when we commit to the variational Bayes approach, each temporal scale is inverted only once. 
       
       In summary, the variational Bayes approach to inverting hierarchical temporal models is a statistical framework that can be used to quantitatively evaluate hypotheses about multiscale data. In particular, it extends a summary statistic approach, in which one would compute summary statistics of the fast variables and then perform \textit{post-hoc} analysis of the slow dynamics. In the following section, we present an example of how this treatment of multiscale dynamical systems has been used successfully. 

       \section{Example: adiabatic dynamic causal modelling}
       We have now introduced three key ingredients for modelling time series data at multiple temporal scales: 1) state-space models, 2) their extension to multiple temporal scales via hierarchical models, and 3) Bayesian inference methods needed to invert these models and evaluate their evidence.
       
       These methods are already proving useful for modelling multi-scale data in practice. For instance, to investigate the role of antibodies against NMDA receptors (NMDAr-Ab) in autoimmune encephalitis, Rosch and colleagues used multiscale modelling to understand how NMDAr-Ab cause abnormal EEG spectra \citep{rosch2018nmda} (see \textbf{Figure~5}). They recorded the local field potential (LFP) of control and NMDAr-Ab-positive mouse models during epileptic seizures induced by pentylenetetrazol (PTZ). The spectra of the LFPs were modelled for short time windows using a canonical microcircuit (CMC) model, composed of several interacting neuronal populations. 
       \begin{figure}
             \centering
             \includegraphics[width=\textwidth]{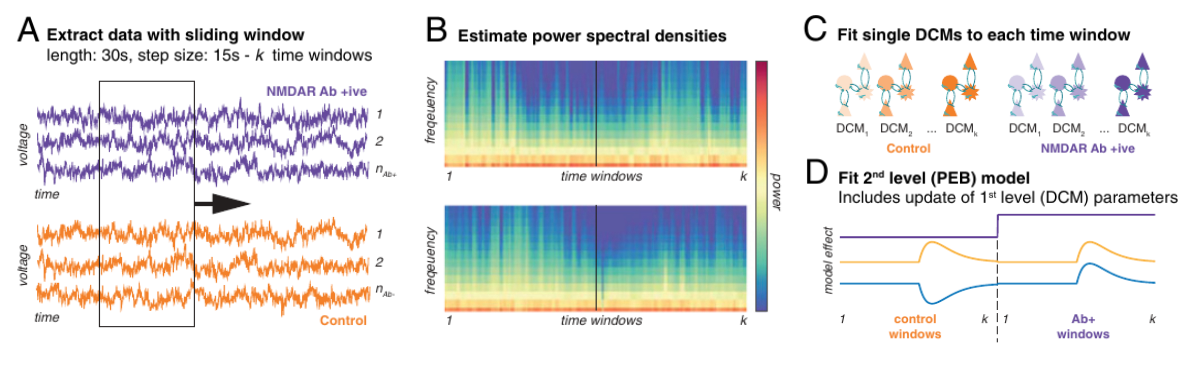}
             \caption{Hierarchical modelling approach used to link slow effects to fast observations. First, the authors extracted sliding window data from their LFP time series (\textbf{A}). Then, they estimated the power spectral density for each window, to produce a time-frequency representation of the data (\textbf{B}). Then they fitted a state-space model, called a Canonical Microcircuit (CMC) dynamic causal model (DCM), for each window of power spectral densities (\textbf{C}). This resulted in a time course of posterior densities for the parameters of the DCM models. Finally, the authors added a second level to the model to test for between-window effects, enabling them to evaluate hypotheses of interest, i.e.\ the interaction between interventions (PTZ concentration, presence or absence of NMDAr-Ab) and the parameters of the CMC model (\textbf{D}). Adapted with permission from \citep{rosch2018nmda}.}
           \label{fig:my_label}
       \end{figure}           

       In this work, the CMC model was the first-level model (encapsulating fast variables) for each time window. The time constants of each neuronal population were the slow variables.  The study evaluated how these slow variables interacted with the PTZ concentration, changing slowly through time, and with the presence of NMDAr-Ab. This was done by comparing models at the second level (i.e., between time windows). Under their model, they observed a strong effect of PTZ on the synaptic time constants of different neuronal populations that was further amplified by the presence of NMDAr-Ab. 
       
       This example illustrates how Bayesian model reduction and hierarchical temporal models can be used in practice. The complex multiscale modelling problem was decomposed in two simpler modelling problems: 
          \begin{enumerate}[i]
              \itemsep0em 
              \item linking slow parameters of the models to the fast measurements,
              \item modelling the effect of slow experimental effects on the slow parameters. 
          \end{enumerate}
        Opting for this decomposition allowed the authors to evaluate their research hypotheses in a principled manner. The method used has been more extensively formalized in the adiabatic DCM framework \citep{Jafarian2021}. The reader interested in an analogous method with discrete slow dynamics can refer to the dynamic effective connectivity framework introduced in \citep{zarghami2020dynamic}.       
   
\section{Concluding remarks and future perspectives}
    The mechanisms of interest for brain research generally span multiple temporal scales. Because of this apparent complexity, here we returned to first principles and pursued established mathematical paths. Evaluating hypotheses amounts to constructing generative models of our observations, for which there are readily available methods. The slaving principle means that slow and fast processes can be analysed separately: we can therefore model each time scale individually, lending a hierarchical structure to our generative models. Moreover, the slaving principle entails switching dynamics, for instance of brain networks dynamics, which licenses the combination of discrete models (e.g., HMMs) for slow changes in brain states and continuous models of fast neuronal dynamics, as we have illustrated.  

    Evaluating and comparing the evidence for hypotheses regarding multiscale time series requires us to invert hierarchical temporal models, i.e. to compute the posterior distribution over model parameters. In most cases, the inversion problem cannot be solved exactly because of both the complexity and hierarchical structure of the model. This forces us to approximate the posterior distribution; a task for which variational Bayes methods are preferred. In particular, variational Bayes can be combined with empirical Bayes to invert arbitrary hierarchical models with empirical priors at intermediate levels. This suggests a simple procedure to invert hierarchical temporal models: first, model the fast temporal scale by fitting a model to each time window of time series data, and then invert these models to obtain a time series of estimated parameters (posteriors), which are expected to change slowly. Finally, use the time series of these slowly changing parameters as observations for the level above in the hierarchy.

    A practical question is how to determine whether it is better to model several time scales, and how to identify which time scales shall be modelled. The general answer is that additional model complexity must be justified by having a better explanation of the data. The trade-off between complexity and accuracy is effectively summarized by the log evidence of the model (and its approximation, the variational free energy).  From this perspective, models accounting for different temporal scales are seen as competing hypotheses, and can be systematically compared using Bayesian model comparison. In other words, modelling different temporal scales is justified if the multiscale model has a greater free-energy than a model without multiple scales.  

    The impact of statistical models for multiscale dynamical systems in neuroscience is poised to grow. A remarkable example is the development of The Virtual Epileptic Patient, which helps construct personalized epilepsy models to aid in clinical comprehension of epileptic cases \citep{jirsa2017virtual,hashemi2020bayesian, jirsa2023personalised}. Several promising directions for developing these models are worth exploring. Firstly, HMMs of EEG and MEG signals could be extended to reflect the itinerancy between the various attractors of the (multistable) system arising from interconnected cortical population, for instance using the escape rates introduced in \citep{cooray2023global,cooray2023modelling}. Such transformation would allow one to relate the switching statistics of the Markov chain to properties of the network, effectively uncovering mechanistic explanations for state transitions.  Secondly, employing models with continuous slow dynamics allows tracking synaptic gain in pyramidal neurons, believed to encode the precision (inverse variance) of top-down predictions on the lower levels of the predictive coding hierarchy. Investigating how these precisions change with experimental manipulations and over time would be particularly interesting. A third avenue for hierarchical models of multiscale time series would be to use theoretical frameworks like Structured Flow on Manifolds (SFM) to model empirical observations. For instance, applying an SFM-based generative model of neuronal activity to motor experiments, in the spirit of \citep{churchland2012neural, saxena2022motor, lara2018different}, could offer valuable insights into the dynamics of motor tasks.

    In conclusion, by combining hierarchical state space models with Bayesian analysis methods, time series data with different temporal scales can be linked and their interactions investigated. This has widespread application within neuroscience research, as well as in empirical science more generally.
        
\section*{Acknowledgments}
    The authors are grateful to Robert Seymour for his useful suggestions.

\section*{Supporting Information}
    The Wellcome Centre for Human Neuroimaging is supported by core funding from Wellcome [203147/Z/16/Z].

\bibliographystyle{apalike} 
\bibliography{references}

\section*{Technical Terms}
\parindent0pt
\textbf{Initial Value Problem} The problem of solving a differential equation with given initial conditions.

\textbf{Response Function} A function that describes the output or behaviour of a system in response to a given input or stimulus.

\textbf{Hidden Markov Model} A probabilistic model that describes a sequence of observable events generated by an underlying unobservable Markov chain.

\textbf{Effective connectivity}The directed influence of neuronal populations on one another.

\textbf{Synergetics} The study of complex systems and their emergent properties, examining interactions between parts to understand their holistic dynamics.

\textbf{Adiabatic Approximation} Neglecting rapidly changing variables in comparison to slowly changing ones.

\textbf{Bayes Factor} A statistical measure that quantifies the evidence supporting one hypothesis, as represented by a model, over another.

\textbf{Posterior Distribution} The updated probability distribution of a variable after considering new data.

\textbf{Bayesian Model Reduction } A statistical technique based on Bayesian inference that simplifies complex models by eliminating irrelevant variables and retaining important ones.

\textbf{Empirical priors} Prior distributions estimated from the empirical distribution of data or lower-level posteriors.

\end{document}